# Verification of TG-61 dose for synchrotron-produced monochromatic x-ray beams using fluence-normalized MCNP5 calculations.

Thomas A. D. Brown, Kenneth R. Hogstrom

*Mary Bird Perkins Cancer Center, 4950 Essen Lane, Baton Rouge, LA 70809 and Department of Physics and Astronomy, Louisiana State University and A & M College, 202 Nicholson Hall, Baton Rouge, LA 70803*

Diane Alvarez, Kenneth L. Matthews II

*Department of Physics and Astronomy, Louisiana State University and A & M College,*
*202 Nicholson Hall, Baton Rouge, LA 70803*

Kyungmin Ham

*Center for Advanced Microstructures and Devices, Louisiana State University and A & M College,*
*6980 Jefferson Highway, Baton Rouge, LA 70806*

**ABSTRACT**

**Purpose:** Ion chamber dosimetry is being used to calibrate dose for cell irradiations designed to investigate photoactivated Auger electron therapy at the Louisiana State University Center for Advanced Microstructures and Devices (CAMD) synchrotron facility. This study performed a dosimetry intercomparison for synchrotron-produced monochromatic x-ray beams at 25 and 35 keV. Ion chamber depth-dose measurements in a polymethylmethacrylate (PMMA) phantom were compared with the product of MCNP5 Monte Carlo calculations of dose per fluence and measured incident fluence.

**Methods:** Monochromatic beams of 25 and 35 keV were generated on the tomography beamline at CAMD. A cylindrical, air-equivalent ion chamber was used to measure the ionization created in a $10 \times 10 \times 10$-cm$^3$ PMMA phantom for depths from 0.6 to 7.7 cm. The American Association of Physicists in Medicine TG-61 protocol was applied to convert measured ionization into dose. Photon fluence was determined using a NaI detector to make scattering measurements of the beam from a thin polyethylene target at angles $30^o$ to $60^o$. Differential Compton and Rayleigh scattering cross sections obtained from *xraylib*, an ANSI C library for x-ray-matter interactions, were applied to derive the incident fluence. MCNP5 simulations of the irradiation geometry provided the dose deposition per photon fluence as a function of depth in the phantom.

**Results:** At 25 keV the fluence-normalized MCNP5 dose overestimated the ion-chamber measured dose by an average of 7.2 ± 3.0% to 2.1 ± 3.0% for PMMA depths from 0.6 to 7.7 cm, respectively. At 35 keV the fluence-normalized MCNP5 dose underestimated the ion-chamber measured dose by an average of 1.0 ± 3.4% to 2.5 ± 3.4%, respectively.

**Conclusions:** These results showed that TG-61 ion chamber dosimetry, used to calibrate dose output for cell irradiations, agreed with fluence-normalized MCNP5 calculations to within approximately 7% and 3% at 25 and 35 keV, respectively.

**I. INTRODUCTION**

Low-energy monochromatic x-rays are being used to investigate photoactivated Auger electron therapy at the Louisiana State University Center for Advanced Microstructures and Devices (CAMD) synchrotron facility. Cell survival studies have been conducted using photoactivation of iododeoxyuridine (IUdR) incorporated into the DNA of the cells. Previous work determined the dependence of Chinese hamster ovary cell survival on IUdR concentration at a beam energy of 35 keV.[1] For that study, the dose output was measured using an air-equivalent ion chamber in a polymethylmethacrylate (PMMA) phantom, by applying the American Association of Physicists in Medicine (AAPM) TG-61 protocol converting measured ionization into dose.[2] The TG-61 protocol was designed to be applied to clinical, polychromatic x-ray beams in the 40 – 300 kV range.[3] In earlier work, GAFCHROMIC® EBT film, calibrated using orthovoltage x-rays, was used to verify ion-chamber measurements in a PMMA phantom at 35 keV.[2] The dose derived from the film was found to underestimate the ion-chamber measured dose by 4.8 ± 2.2% at a depth of 2 cm. Also, an x-ray scatter method was used as a tool for a fluence-based dose comparison.[4] NaI detector measurements of x-rays scattered from a thin polyethylene target at angles $15^o$ to $60^o$, were used to determine the beam fluence by applying the Klein-Nishina differential cross section for Compton scattering of polarized photons. The calculated fluence was used to normalize a MCNP5-calculated depth-dose profile for a PMMA phantom, which was compared with the ion-chamber measured dose. The fluence-based dose overestimated that from the ion chamber by an average of 6.4 ± 0.8% for PMMA depths from 0.6 to 6.1 cm.[4] In applying this technique to a 25 keV beam, two deficiencies in the methodology of Dugas *et al.*[4] were identified: (1) the scattering cross section used for the fluence calculations was calculated assuming that electron binding effects on Compton scattering could be ignored, and (2) it was assumed that photons measured by the NaI detector (after background subtraction) were Compton scattered from the target, i.e. Rayleigh scattering was ignored. While these assumptions are reasonably valid for higher energies (above ~ 200 keV), they are not realistic at 35 keV and below.

Incoherent scatter factors available in the literature[5,6] indicate that electron binding effects are significant for the target material and angular range used by Dugas *et al.*[4] At 35 keV differential Compton cross sections obtained from *xraylib*[7], an ANSI C library for x-ray-matter interaction data, show that the pre-collision momentum of the electrons in the target effectively reduces the Compton cross section by 26% to 3% for angles $15^o$ to $60^o$, respectively. Hence, for the four measurement angles averaged by Dugas ($15^o$, $30^o$, $45^o$, and $60^o$), the Compton cross section was over predicted by 12%. FIG.

1 shows a comparison of the differential Compton cross section per gram molecular weight calculated with and without electron binding effects at 35 and 25 keV.

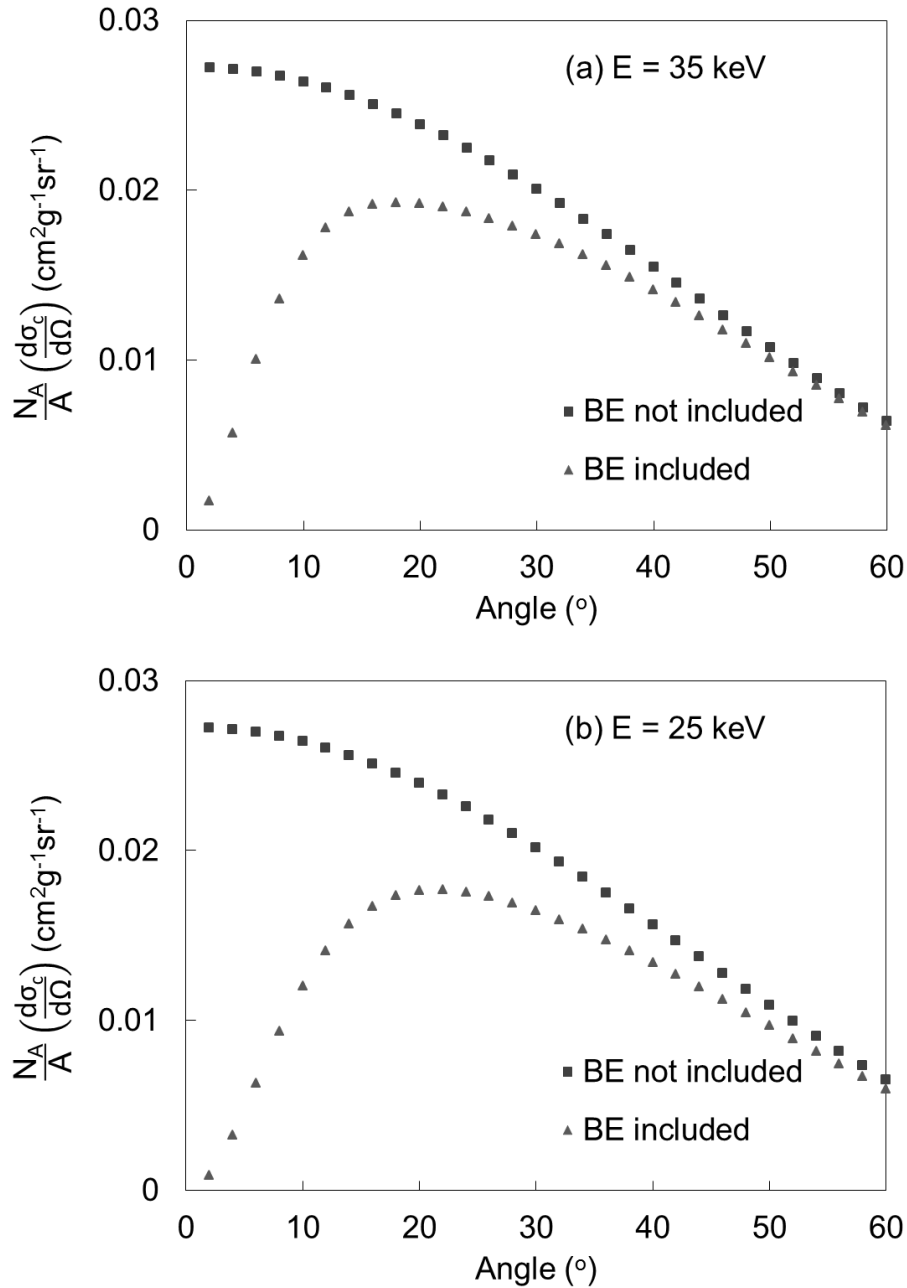

FIG 1: Differential Compton cross section per gram molecular weight versus scattering angle for polyethylene at (a) 35 keV and (b) 25 keV. Squares: differential cross section calculated using the formulism described by Dugas et al.[4] which ignores electron binding effects (BE not included). Triangles: differential cross section values obtained from *xraylib*[7] which includes electron binding effects (BE included). Cross section values were determined for the case of x-rays 100% polarized in the plane of the synchrotron.

Calculations using *xraylib* have also shown that the Rayleigh scatter contribution from polyethylene at 35 keV is significant. The ratio of the differential Rayleigh cross section to the differential Compton cross section varies between 0.74 and 0.07 for angles $15^o$ to $60^o$, respectively. Hence, for the four measurement angles averaged by Dugas, the Rayleigh scattered events accounted for 21% of the events scattered from the target. FIG. 2 illustrates the variation of the differential Rayleigh and Compton cross sections per gram molecular weight as a function of angle for 35 and 25 keV.

The inclusion of Rayleigh scatter and electron binding effects partially offset each other at 35 keV, although these effects become more prominent at lower energies. FIG. 3 shows a comparison of the total differential scattering cross section (Rayleigh plus Compton including electron binding effects) and the differential Compton scattering cross section ignoring electron binding effects for 35 and 25 keV. Including the contribution from Rayleigh scatter and the electron binding effects increases the scattering cross section by approximately 30%, 11%, 7%, and 4% at $15^o$, $30^o$, $45^o$, and $60^o$, respectively. Hence, the scattering cross section used by Dugas was underestimated by an average of 13%. Given that the fluence-normalized MCNP5 dose was reported as overestimating the ion-chamber measured dose by $6.4 \pm 0.8\%$, applying a cross section correction of 13% yields a fluence-normalized MCNP5 dose that *underestimates* the ion-chamber measured dose by $5.8 \pm 0.8\%$, since a larger cross section reduces the magnitude of the fluence calculated from the number of x-rays scattered into the detector.

The CAMD Auger electron therapy studies are now investigating the dependence of rat glioma cell survival for beam energies above and below the iodine K-edge (33.2 keV). Although the accuracy of the ion-chamber dose was studied at 35 keV[2,4], this paper reports the ion-chamber dose verification at 25 and 35 keV, by applying a more rigorous version of the fluence analysis used by Dugas *et al.* to a new set of x-ray scatter measurements.

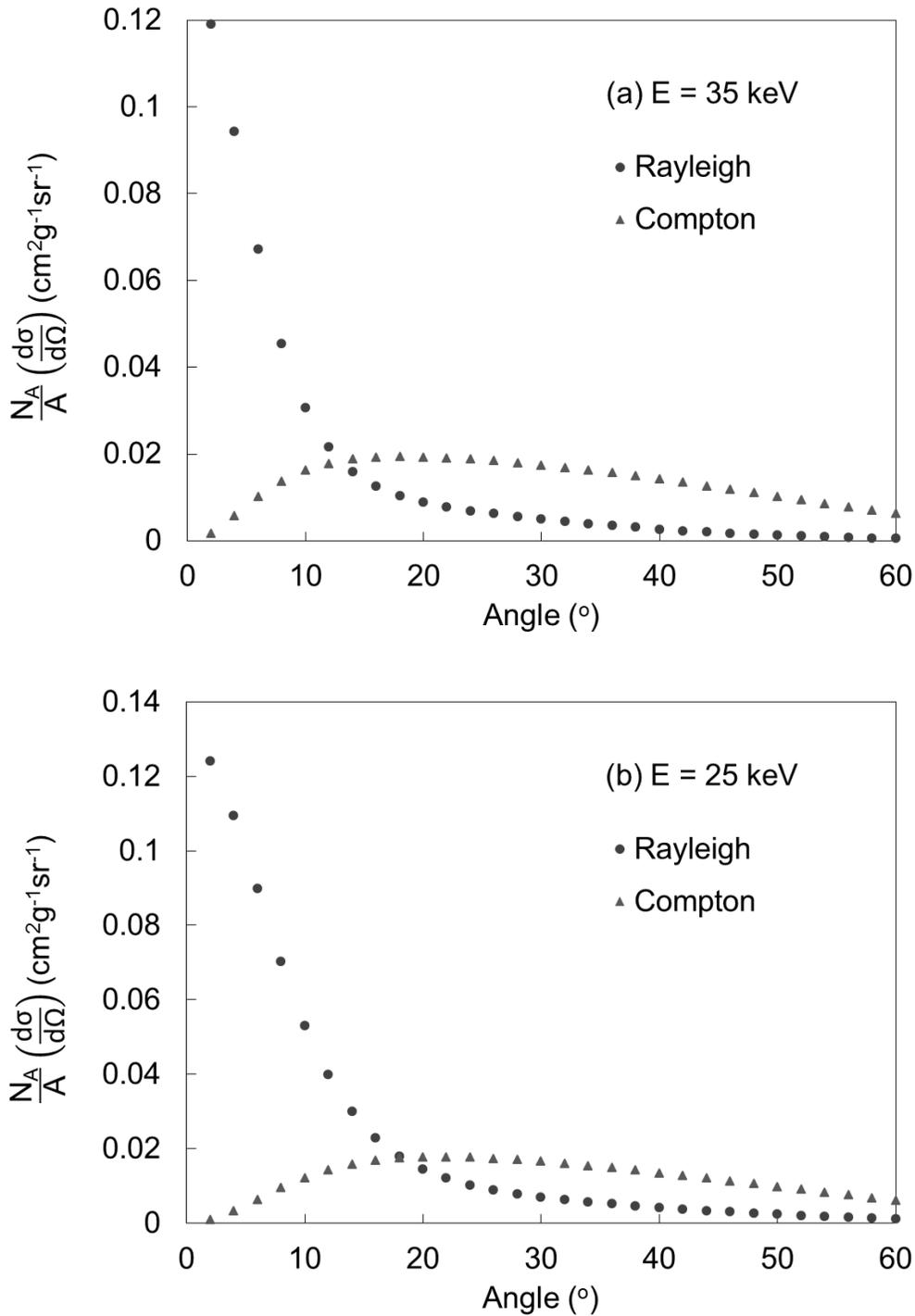

FIG. 2: Differential cross section per gram molecular weight versus scattering angle for Rayleigh and Compton scattering from polyethylene at (a) 35 keV and (b) 25 keV. Cross section values were obtained from *xraylib* for the case of x-rays 100% polarized in the plane of the synchrotron.

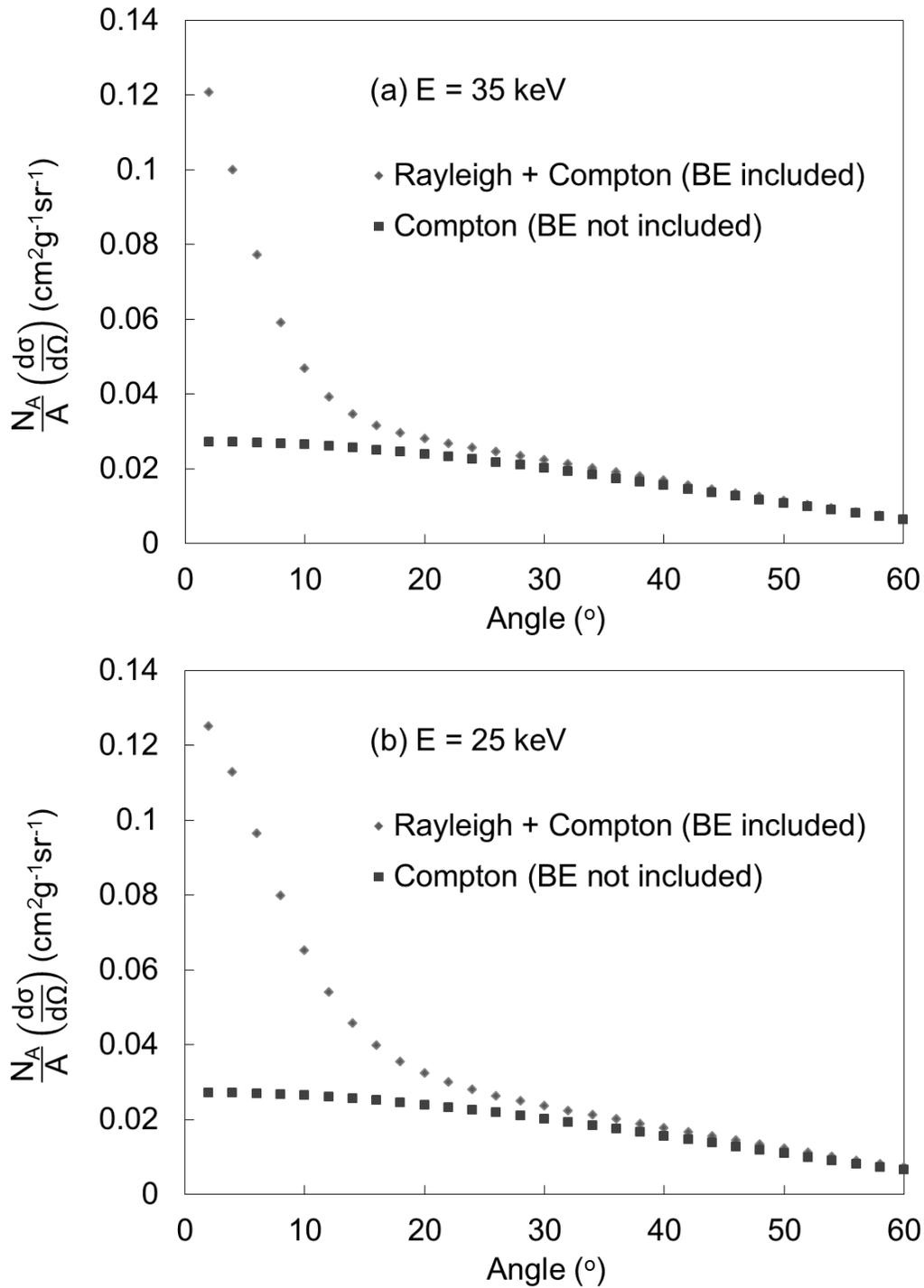

FIG. 3: Total and Compton differential cross sections per gram molecular weight versus scattering angle for polyethylene at (a) 35 keV and (b) 25 keV. Diamonds: total differential cross section obtained using Rayleigh and Compton cross section values obtained from *xraylib*[7] which include electron binding effects (BE included). Squares: differential Compton cross section calculated using the formulism described by Dugas et al.[4] which ignores electron binding effects (BE not included). Cross section values were determined for the case of x-rays 100% polarized in the plane of the synchrotron.

## II. METHODS AND MATERIALS

### II.A Irradiation source

Monochromatic x-ray beams of 25 and 35 keV were generated on the tomography beamline at CAMD. A 1.3-GeV electron beam ($I_{max}$ = 220 mA) was transported through a three-pole superconducting wiggler magnet ($B_{max}$=7T), creating a polychromatic beam. Monochromatic x-rays ($\Delta E/E \approx 2$ %) were selected by transporting the beam through a W-B$_4$C double-multilayer monochromator (Oxford Danfysik, UK). Due to size restrictions imposed by the monochromator and beamline slits, the resulting monochromatic beam was approximately 3.0-cm wide × 0.2-cm high. The narrow beam was filtered using 640 μm Al since low-energy x-ray contamination can be significant. The energy of the beam was verified by measuring Debye-Scherrer cones produced from Si640c powder diffraction[4]. A flat-panel XRD 0820 CN3 detector (PerkinElmer, Waltham, MA) measured the resulting diffraction rings, allowing for energy precision to within 0.1 keV. An effective broad beam approximately 3.0-cm wide × 2.5-cm high was created by vertically oscillating the irradiation target through the path of the fixed narrow beam at 0.125 cms$^{-1}$ (40 s period). Target oscillation was achieved using a screw-drive motion stage (Velmex, Inc., Bloomfield, NY) controlled by a user-programmed LabVIEW (National Instruments Corporation, Austin, TX) interface. Previous measurements have shown that the effective broad beam can be considered parallel.[4]

### II.B Ion chamber dosimetry

The dose delivered by the beam in a 10×10×10-cm$^3$ PMMA phantom was measured using a calibrated 0.23-cm$^3$ Scanditronix Wellhofer model FC23-C cylindrical, air-equivalent ion chamber (Scanditronix Wellhofer GmbH, Schwarzenbruck, Germany) with a Modified Keithley 614 Electrometer (CNMC Company, Best Medical, Nashville, TN). The ion chamber was used to measure the ionization created by the effective broad beam along its central axis at PMMA depths from 0.6 to 7.7 cm. The length of a broad beam irradiation was specified in terms of the number of complete stage oscillations, ensuring that the dose delivery was uniform in the vertical direction. Each irradiation measurement was conducted for 320 s, corresponding to eight complete stage oscillations. The x-ray dose output (dose per unit time) was proportional to the synchrotron storage ring current, which slowly decayed over the time between electron injections into the ring (~ 7 hours). Using the average ring current for each irradiation, the measured ionization was normalized to a ring current of 100 mA.

The AAPM TG-61 protocol[3] for determining dose to water ($D_w$) for medium energy x-rays (100 kV – 300 kV) at 2-cm depth, was applied to convert the normalized ionization ($M_{norm}$) at all depths into dose:

$$D_w = M_{norm} P_{elec} P_{TP} P_{ion} P_{pol} P_{Q,cham} N_k \left(\frac{\mu_{en}}{\rho}\right)_{AIR}^{WATER}, \qquad (1)$$

where $P_{elec}$ is the electrometer accuracy correction factor, $P_{TP}$ is the ambient temperature and pressure correction factor, $P_{ion}$ is the ion recombination correction factor, $P_{pol}$ is the polarity effect correction factor, $P_{Q,cham}$ is the overall chamber correction factor, $N_k$ is the air-kerma calibration factor, and $\left(\frac{\mu_{en}}{\rho}\right)_{Air}^{WATER}$ is the ratio of the water-to-air mass-energy absorption coefficients. The ion chamber correction and calibration factors were obtained in the same way as described by Oves *et al.*[2] and are shown for both energies in Table 1. The ion chamber measurements used to calculate $P_{ion}$ and $P_{pol}$ were conducted at a PMMA depth of 0.6 cm using the same broad beam geometry as the depth-dose measurements. Irradiations were typically performed for 160 s (four stage oscillations), and the measured ionization was normalized to a ring current of 100 mA. $P_{ion}$ was determined for the case of a continuous beam using high and low electrometer bias voltages of -300 and -150 V, respectively. Values for $P_{Q,cham}$ were difficult to determine since the energies and field size used for these measurements lay outside the range of data available for this correction factor in TG-61. Estimates of $P_{Q,cham}$ were obtained by using $P_{Q,cham} = 0.995$ for the similar NE2611/NE2561 chambers and for a 0.1 mm Cu HVL beam in TG-61 Table VIII, and then applying a field size correction factor of 1.005 by extrapolating data in TG-61 Figure 4 for the broad beam size (7.5 cm$^2$) used in this work. $N_k$ was determined using a linear fit to ADCL calibrated values measured for a 120 kVp beam (HVL=6.96 mm Al) and an 80 kVp beam (HVL = 2.96 mm Al), which were $1.215 \times 10^8$ Gy C$^{-1}$ and $1.219 \times 10^8$ Gy C$^{-1}$, respectively. The HVL values were used to interpolate and extrapolate $N_k$ values at 35 keV (HVL = 3.33 mm Al) and 25 keV (HVL = 1.12 mm Al). Mass-energy absorption coefficients were interpolated for 25 and 35 keV from NIST tables[8] and used to calculate values for $\left(\frac{\mu_{en}}{\rho}\right)^{WATER}_{AIR}$ at both energies.

| TG-61 calibration factor | 25 keV | 35 keV |
|---|---|---|
| $P_{elec}$ | 0.987 | 0.987 |
| $P_{TP}$ | 1.006 – 1.016 | 1.009 – 1.018 |
| $P_{ion}$ | 0.999 – 1.002 | 0.995 – 1.000 |
| $P_{pol}$ | 0.999 – 1.001 | 1.003 – 1.009 |
| $P_{Q,cham}$ | 1.000 | 1.000 |
| $N_k$ | $1.221 \times 10^8$ Gy C$^{-1}$ | $1.219 \times 10^8$ Gy C$^{-1}$ |
| $\left(\frac{\mu_{en}}{\rho}\right)^{WATER}_{AIR}$ | 1.019 | 1.015 |

TABLE 1: TG-61 ion chamber calibration and correction factors used for dose calculations at 25 and 35 keV. Measurements of $P_{TP}$, $P_{ion}$ and $P_{pol}$ were repeated for each set of depth-dose measurements, and the range of values obtained are shown here.

The principle source of uncertainty in the normalized ionization values arose from small variations in the beam output that were independent of the ring current. These variations can arise as a result of changes in the phase space of the ring electrons, beamline vacuum fluctuations, or beam heating of the monochromator. The standard deviation of multiple normalized ionization values measured at a single PMMA depth was used to estimate the size of this uncertainty. The total uncertainty in the corrected, normalized, ionization value ($M_{norm}P_{elec}P_{TP}P_{ion}P_{pol}P_{Q,cham}$) used to determine the TG-61 dose was found by propagating the uncertainty in $P_{ion}$, $P_{pol}$, and $M_{norm}$, and was determined to be ± 3%.

**II.C Fluence measurements**

X-ray scattering measurements of the fixed narrow beam were used to determine the fluence on the central axis of the beam. A similar experimental setup to that described by Dugas et al.[4] was utilized for these measurements. The $3.0 \times 0.2$ cm$^2$ narrow beam was collimated horizontally to be approximately $0.1 \times 0.2$ cm$^2$ using 0.24-cm thick tungsten plates. The collimated beam was incident on a 0.05-cm thick polyethylene foil, which scattered into a lead-shielded 0.1-cm thick × 2.54-cm diameter NaI(Tl) scintillator detector (Alpha Spectra, Grand Junction, CO). The energy spectra for photons scattered from the foil at 15° intervals from 15° to 60° with respect to the beam axis was recorded. The target foil remained parallel to the front face of the NaI detector for all angles. The scattered beam was collimated at the face of the detector using a 0.17-cm thick lead disc with either a 0.123 or 0.181-cm$^2$ rectangular aperture to reduce the event rate in the data acquisition, and also to provide a well-defined solid angle for the scattered x-rays. The target-aperture distance was 16.1 cm. Pulses from the detector were amplified and then digitized using an ORTEC Model 926 Multichannel Buffer (ORTEC, Oak Ridge, TN), which generated an 8191-channel energy spectrum. The data acquisition system was turned on at least 2 hours prior to recording data, so that any electronic gain changes associated with the warm-up of the system were minimized. Spectra were typically acquired for 250 s or less at each angle. For each scattering angle a measurement was made with and without the target, so that x-rays scattering into the detector from objects other than the target could be subtracted. A background measurement (~ 700 s) with no beam, acquired for each measurement condition, was found to make an insignificant contribution to the number of detected events. Measurements made at 15° were subsequently rejected due to the relatively large data acquisition dead times observed at both energies. The dead time at 15° was typically a factor of two larger compared to the other angles, and was as high as 40% at 25 keV.

**II.D Fluence calculation**

The central axis beam fluence was determined from the number of photon counts in the background subtracted x-ray spectra. Two sets of photon events could be identified in the spectra: those photons that were transmitted through the monochromator at the desired energy, satisfying the Bragg condition ($n\lambda = 2d\sin\theta$) for n=1, and those photons that were transmitted satisfying the condition for n=2 ($E=2E_{Beam}$). Most of the detected events were n=1 photons. These photons exhibited a well-defined peak in all of the spectra as a result of Compton and Rayleigh scatter from the target. The energy resolution of the detector ($\Delta E/E \approx 20\,\%$) did not allow for discrimination between the two types of scatter. The resolution was poor but not unusual for a NaI detector. An iodine-escape peak was also evident in the spectra for those measurements made at 35 keV. At 25 keV less than 4% of the detected events corresponded to n=2 photons and formed a low broad peak in the spectra at high energy. At 35 keV the n=2 photons accounted for less than 2% of the number of detected events and were ignored in the analysis used by Dugas et al.[4] However, they have been included in the present analysis for both energies. FIG. 4 shows examples of the background-subtracted energy spectra recorded at 25 and 35 keV. An energy calibration for the spectra was obtained using an $^{55}$Fe source (5.9 keV), and by calculating the energy of the scattered x-rays from the measured incident energy by assuming that all of the x-rays were Compton scattered at angles above 45°.

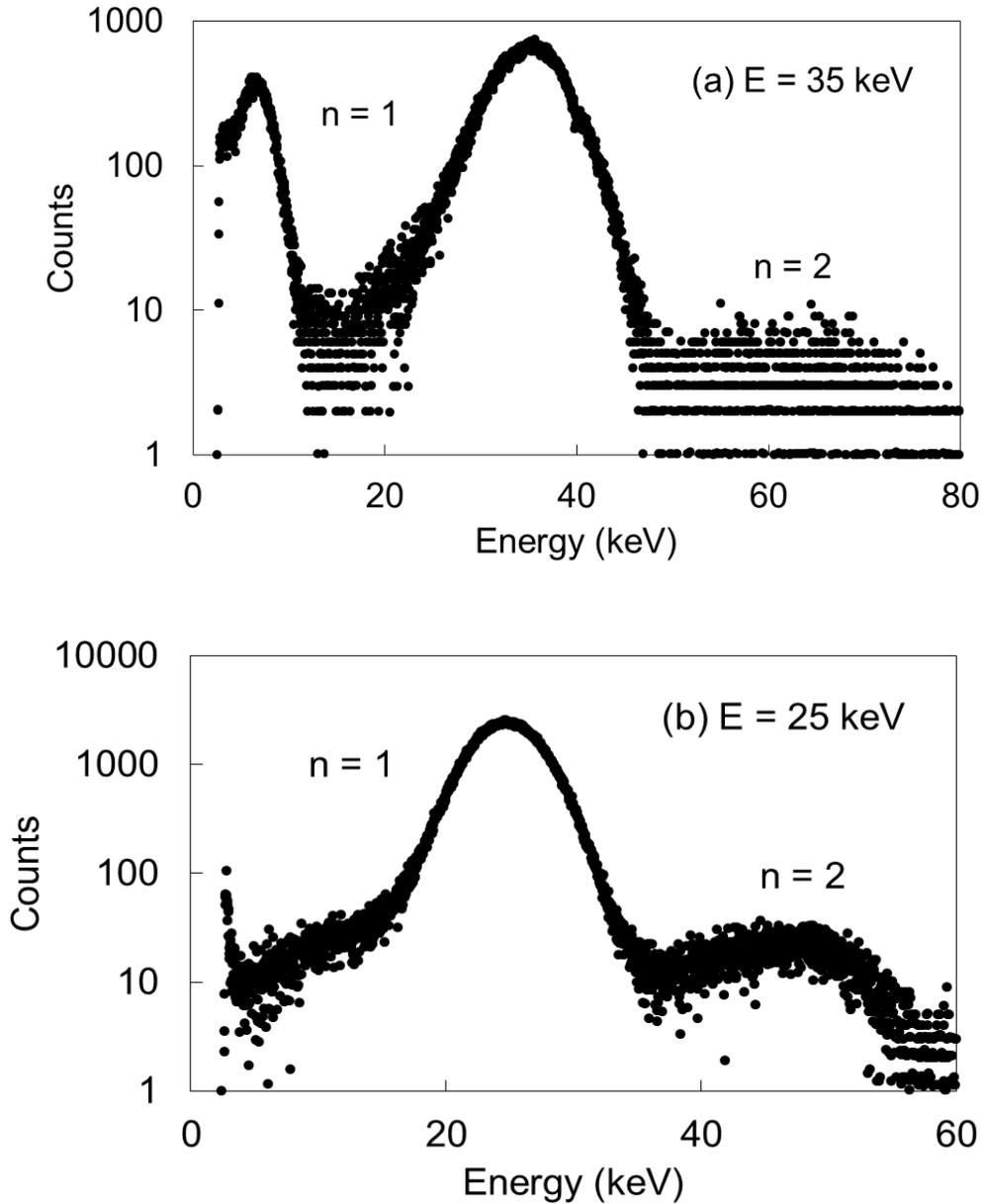

FIG 4: Background-subtracted energy spectra measured at (a) 35 keV for θ = 45° and (b) 25 keV for θ = 60°. Spectra were acquired for live times of 170 and 135 s, respectively. In spectrum (a) the peaks at ~ 35 keV and ~ 6 keV (iodine escape peak) correspond to n=1 photons which have been Compton or Rayleigh scattered from the target. The events at high energy correspond to n=2 photons. In spectrum (b) the peaks at ~ 25 keV and ~ 50 keV correspond to n=1 and n=2 photons which have been Compton or Rayleigh scattered from the target.

The fluence contribution from the n=1 and n=2 photons were calculated separately. The count rate $\dot{C}$ for each type of photon was obtained by summing the number of counts in the respective peaks and dividing by the data acquisition live time. The incident beam fluence rate, $\dot{\emptyset}$, normalized to a storage ring current of 100 mA using the average ring current, $I$, during data acquisition, was determined by applying the following equation:

$$\dot{\emptyset} = \frac{\dot{C}}{\left(\frac{N_A}{A}\left(\frac{d\sigma_T}{d\Omega}\right) \times \rho t \times \Delta\Omega \times A_{\text{Beam}} \times \epsilon\right)} \times \frac{100 \text{ } mA}{I}, \quad (2)$$

where $\frac{N_A}{A}\left(\frac{d\sigma_T}{d\Omega}\right)$ is the total differential scattering cross section per gram molecular weight (cm$^2$g$^{-1}$sr$^{-1}$), $\rho t$ is the target thickness (gcm$^{-2}$), $\Delta\Omega$ is the solid angle subtended by the detector collimator aperture (sr), $A_{\text{Beam}}$ is the cross-sectional area of the incident beam (cm$^2$), and $\epsilon$ is the intrinsic efficiency of the detector.[4] The cross-sectional area of the beam – width × effective height – was determined by exposing GAFCHROMIC® EBT2 film (International Specialty Products, Wayne, NJ) to the collimated narrow beam. The resulting beam spot image was digitized using an Epson 1680 Professional flatbed scanner (Seiko Epson Corporation, Nagano, Japan) and measured using ImageJ v1.42q (National Institutes of Health, Bethesda, MD). The intrinsic efficiency as a function of energy for a 0.1-cm-thick NaI crystal was obtained from Knoll[9]. The total differential scattering cross section per gram molecular weight was defined as the sum of the Compton and Rayleigh differential cross sections per gram molecular weight, obtained for polyethylene from *xraylib*[7]:

$$\frac{N_A}{A}\left(\frac{d\sigma_T}{d\Omega}\right) = \frac{N_A}{A}\left(\frac{d\sigma_C}{d\Omega}\right) + \frac{N_A}{A}\left(\frac{d\sigma_R}{d\Omega}\right). \quad (3)$$

The differential cross section for scattering from an atom of atomic number Z ($cm^2 atom^{-1} sr^{-1}$) is characterized in terms of an atomic form factor $F(E, \theta, Z)$ and an incoherent scattering factor $S(E, \theta, Z)$ for Rayleigh and Compton scattering, respectively:

$$\frac{d\sigma_R}{d\Omega} = r_e^2 (1 - \sin^2\theta \cos^2\varphi) F^2(E, \theta, Z), \tag{4}$$

$$\frac{d\sigma_C}{d\Omega} = \frac{r_e^2}{2} \left(\frac{E'}{E}\right)^2 \left(\frac{E}{E'} + \frac{E'}{E} - 2\sin^2\theta \cos^2\varphi\right) S(E, \theta, Z), \tag{5}$$

where $r_e$ is the classical electron radius ($2.82 \times 10^{-13}$ cm), $\theta$ is the scattering angle, $\varphi$ is the polarization angle, $E$ is the incident photon energy, and $E'$ is the scattered photon energy at angle $\theta$.[7] For the compound polyethylene $(C_2H_4)_n$, the differential cross sections per gram molecular weight can be expressed in terms of the constituent atomic cross sections:

$$\left[\frac{N_A}{A}\left(\frac{d\sigma_R}{d\Omega}\right)\right]_{C_2H_4} = \frac{2A_C}{A_{C_2H_4}} \left[\frac{N_A}{A}\left(\frac{d\sigma_R}{d\Omega}\right)\right]_C + \frac{4A_H}{A_{C_2H_4}} \left[\frac{N_A}{A}\left(\frac{d\sigma_R}{d\Omega}\right)\right]_H, \tag{6}$$

$$\left[\frac{N_A}{A}\left(\frac{d\sigma_C}{d\Omega}\right)\right]_{C_2H_4} = \frac{2A_C}{A_{C_2H_4}} \left[\frac{N_A}{A}\left(\frac{d\sigma_C}{d\Omega}\right)\right]_C + \frac{4A_H}{A_{C_2H_4}} \left[\frac{N_A}{A}\left(\frac{d\sigma_C}{d\Omega}\right)\right]_H, \tag{7}$$

where $A_C$ and $A_H$ are the atomic weights of carbon and hydrogen, and $A_{C_2H_4}$ is the molecular weight of polyethylene. For a user-defined compound, $E$, $\theta$, and $\varphi$, *xraylib* calculates the Rayleigh cross section using atomic form factor values taken from Hubbell *et al.*[5], and determines the Compton cross section using incoherent scatter factor values taken from Cullen *et al.*[6] Previous measurements have shown that the x-rays are polarized in the plane of the synchrotron ($\varphi = 0$).[4] Table 2 shows examples of the scattering factors and differential cross sections per gram molecular weight obtained for polyethylene for E = 25 and 50 keV and E = 35 and 70 keV .

| E (keV) | θ (°) | Carbon F(E, θ, Z) | Hydrogen F(E, θ, Z) | $\left[\frac{N_A}{A}\left(\frac{d\sigma_R}{d\Omega}\right)\right]_{C_2H_4}$ (cm²g⁻¹sr⁻¹) | Carbon S(E, θ, Z) | Hydrogen S(E, θ, Z) | $\left[\frac{N_A}{A}\left(\frac{d\sigma_R}{d\Omega}\right)\right]_{C_2H_4}$ (cm²g⁻¹sr⁻¹) |
|---|---|---|---|---|---|---|---|
| 25 (n=1) | 30 | 1.648 | 0.062 | 0.0070 | 4.526 | 0.996 | 0.0165 |
|  | 45 | 1.351 | 0.017 | 0.0031 | 5.002 | 1.000 | 0.0116 |
|  | 60 | 1.105 | 0.007 | 0.0010 | 5.357 | 1.000 | 0.0060 |
| 50 (n=2) | 30 | 1.068 | 0.006 | 0.0029 | 5.402 | 1.000 | 0.0184 |
|  | 45 | 0.627 | 0.001 | 0.0007 | 5.800 | 1.000 | 0.0126 |
|  | 60 | 0.366 | 0.0005 | 0.0001 | 5.932 | 1.000 | 0.0062 |
| 35 (n=1) | 30 | 1.394 | 0.021 | 0.0050 | 4.931 | 1.000 | 0.0174 |
|  | 45 | 1.031 | 0.005 | 0.0018 | 5.445 | 1.000 | 0.0122 |
|  | 60 | 0.727 | 0.002 | 0.0004 | 5.730 | 1.000 | 0.0062 |
| 70 (n=2) | 30 | 0.687 | 0.002 | 0.0012 | 5.759 | 1.000 | 0.0191 |
|  | 45 | 0.312 | 0.0004 | 0.0002 | 5.951 | 1.000 | 0.0125 |
|  | 60 | 0.154 | 0.0001 | 0.00002 | 5.988 | 1.000 | 0.0060 |

TABLE 2: Scatter factors and differential cross sections per gram molecular weight for polyethylene for E = 25 and 50 keV and E = 35 and 70 keV. The values shown were obtained from *xraylib*.[7]

Differential cross sections per gram molecular weight were obtained for n=1 and n=2 photons using the energy value obtained from the Si640c powder-diffraction measurements. By applying Equation (2) the incident fluence rate for both types of photon was calculated for each scattering angle. An average incident fluence rate for n=1 and n=2 photons was obtained by averaging the results at angles 30°, 45°, and 60°. The broad beam fluence, $\emptyset$, for an exposure time of 320 s was calculated by applying the following equation:

$$\emptyset = \dot{\emptyset} \times \frac{h}{H} \times 320 \, s, \tag{8}$$

where $h$ is the effective height of the narrow beam (cm) and $H$ is the height of the broad beam (2.5 cm).

There were four principle sources of error associated with the broad beam fluence: (1) counting statistics, (2) beam output fluctuations independent of the storage ring current, (3) the uncertainty in the solid angle subtended by the detector collimator aperture, and (4) the uncertainty in the width of the collimated narrow beam. By substituting Equation (2) into Equation (8) it can be seen that the broad beam fluence was independent of the height of the narrow beam. The first two sources of error were responsible for the spread of the fluence rate values obtained at angles 30° to 60°, and their combined effect was determined by calculating the standard error in the mean. For n=1 events the standard error was typically less than ±2% of the mean. For n=2 events the standard error was approximately ±20%. The photon flux from the wiggler magnet falls off rapidly above 35 keV and this high-energy region of the photon spectrum, beyond the energy range utilized for measurement on the synchrotron, may be relatively unstable. Notwithstanding this large uncertainty, the low number of n=2 events resulted in a small n=2 contribution to the total uncertainty in the final dose. The last two sources of error were systematic errors associated with the setup of the experimental apparatus and the measurement technique used to determine the cross-sectional area. The solid angle and beam width uncertainties were estimated to be approximately ±2.5% and ±2%, respectively.

**II.E Monte Carlo simulations**

The transport of a 3.0×2.5-cm$^2$ x-ray beam through a homogeneous PMMA phantom was simulated using the General Monte Carlo N-Particle Transport Code, MCNP5 (Los Alamos National Laboratory, Los Alamos, NM). The simulations were run in photon and electron transport mode only. Irradiation geometry in MCNP5 was modeled using monochromatic photons of measured beam energy originating from a uniform 3.0×2.5-cm² distribution located 10-cm upstream of a solid PMMA block measuring 10×10×12.5 cm³. Photons traveled along parallel trajectories from the source toward the phantom surface. Dose deposition per photon fluence was determined in 0.1×0.1×0.1-cm³ voxels along the phantom's central axis using the F6 cell heating tally for photons. Event histories ranging from 2×10⁷ to 6×10⁷ were obtained for the simulations, yielding a statistical uncertainty in the dose deposition per photon fluence that increased from <0.6% to <1.6% for depths 0.6 to 7.7 cm. For each set of fluence measurements, simulations were performed for n=1 and n=2 photons using the energy value obtained from the Si640c powder-diffraction measurements.

MCNP5-calculated values of dose per fluence were converted to dose to water using the measured broad beam fluence and the $\left(\frac{\mu_{en}}{\rho}\right)_{PMMA}^{WATER}$ ratio derived from NIST values[8]. The two resulting depth-dose profiles were summed to produce a total depth-dose profile $D(z)$:

$$D(z) = \emptyset_{n=1} \frac{D}{\emptyset}(E_{Beam}, z) + \emptyset_{n=2} \frac{D}{\emptyset}(2E_{Beam}, z), \qquad (9)$$

where z is the phantom depth and $\frac{D}{\emptyset}$ is the dose per fluence calculated by MCNP5. The total depth-dose profile was used for comparison with the ion-chamber measured depth-dose profile.

The total uncertainty in the fluence-normalized MCNP5 doses was determined by propagating the statistical uncertainty arising from the MCNP5 simulations and the uncertainties associated with the fluence measurements discussed in Section II D. The total uncertainty was dominated by the fluence measurement uncertainties and was determined to be approximately ±4%.

## III. RESULTS AND DISCUSSION

Three sets of intercomparison measurements were performed at 25 keV, and two sets were performed at 35 keV. FIG. 5 shows an example at 35 and 25 keV comparing the depth-dose profiles constructed from the ion chamber measurements and the fluence-normalized MCNP5 calculations. Doses were compared for a 100 mA exposure time of 320 s.

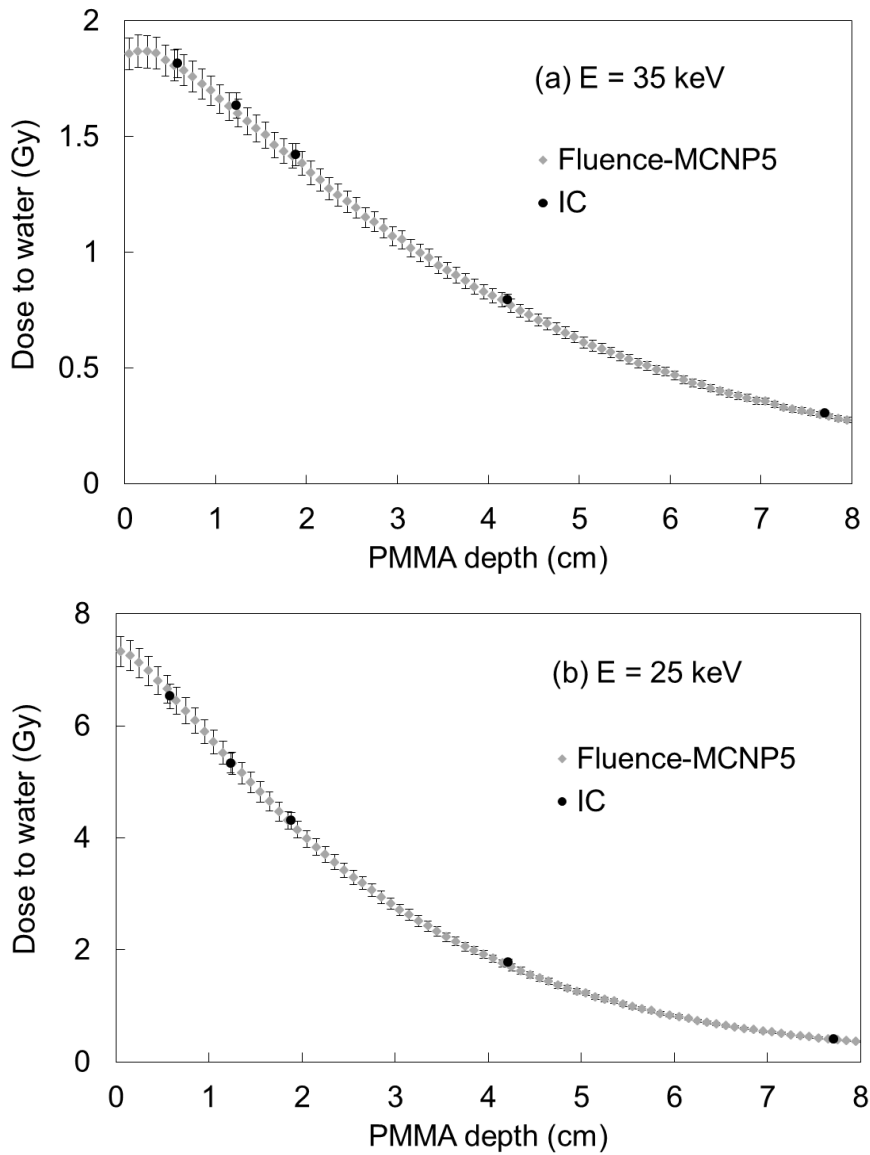

FIG. 5: Dose to water versus PMMA depth at (a) 35 keV (1 of 2 data sets) and (b) 25 keV (1 of 3 data sets). Ion chamber measurements are compared with the product of a MCNP5 calculation and measured beam fluence for an exposure time of 320 s. Dose values were normalized to a synchrotron storage ring current of 100 mA.

Table 3 shows the % difference between the dose measured using the ion chamber and the dose determined from the fluence-normalized MCNP5 calculations for PMMA depths from 0.6 to 7.7 cm. Weighted ($1/\sigma^2$) average values for the % difference have been calculated for each nominal energy setting. At 25 keV the fluence-normalized MCNP5 dose overestimated the ion-chamber measured dose by an average of 7.2 ± 3.0% to 2.1 ± 3.0% for PMMA depths from 0.6 to 7.7 cm, respectively. The difference between the ion-chamber dose and the fluence-MCNP5 dose decreased by ≈ 5% with depth for each set of measurements. This effect is partly due to the absence of the iodine-escape contribution in the n=2 fluence calculations. Although the n=1 escape peak was clearly evident in the spectra (see FIG. 4), the escape peak for n=2 events was buried underneath the n=1 scatter peak, and was included as part of the n=1 fluence. However, by using the n=1 escape fraction to estimate the n=2 escape fluence contribution, the resulting change in the average % difference is < 0.5% for all depths.

| $E_{mono}$ (keV) | $E_{meas}$ (keV) | % difference: $\left(\frac{(D_{IC} - D_{\emptyset,MCNP5})}{D_{IC}}\right) \times 100\%$ | | | | |
|---|---|---|---|---|---|---|
| | | z = 0.6 cm | z = 1.2 cm | z = 1.9 cm | z = 4.2 cm | z = 7.7 cm |
| 25 | 25.3 | -0.9 ± 5.0 | -0.6 ± 5.0 | 1.1 ± 4.9 | 3.4 ± 4.9 | 3.8 ± 5.0 |
| 25 | 25.4 | -11.4 ± 4.9 | -10.5 ± 4.9 | -11.1 ± 4.9 | -10.6 ± 4.9 | -7.2 ± 5.0 |
| 25 | 25.5 | -9.7 ± 5.8 | -8.5 ± 5.7 | -7.7 ± 5.7 | -5.6 ± 5.6 | -3.1 ± 5.6 |
| **Average at 25 keV** | | **-7.2 ± 3.0** | **-6.4 ± 3.0** | **-5.8 ± 3.0** | **-4.1 ± 2.9** | **-2.1 ± 3.0** |
| 35 | 35.3 | 1.7 ± 4.6 | 2.7 ± 4.6 | 2.1 ± 4.6 | 2.8 ± 4.6 | 4.6 ± 4.6 |
| 35 | 35.9 | 0.2 ± 5.0 | -1.8 ± 5.1 | 0.3 ± 5.0 | -2.4 ± 5.2 | -0.02 ± 5.1 |
| **Average at 35 keV** | | **1.0 ± 3.4** | **0.7 ± 3.4** | **1.3 ± 3.4** | **0.5 ± 3.4** | **2.5 ± 3.4** |

TABLE 3: % difference between the ion chamber and the fluence-normalized MCNP5 doses determined at PMMA depths 0.6 to 7.7 cm. $E_{mono}$ is the energy setting of the monochromator, and $E_{meas}$ is the measured energy value obtained from the Si640c powder-diffraction measurements. Weighted average values for the % difference have been calculated for each nominal energy setting.

At 35 keV the fluence-normalized MCNP5 dose underestimated the ion-chamber measured dose by an average of 1.0 ± 3.4% to 2.5 ± 3.4% for PMMA depths from 0.6 to 7.7 cm, respectively. In contrast with the 25 keV results, the data showed no significant trend with depth. The 35 keV results agree with the dosimetry intercomparison work reported by Oves *et al.*, where the dose derived from GAFCHROMIC® EBT film was found to underestimate the ion-chamber dose by 4.8 ± 2.2% at a depth of 2 cm.[2] These results are also consistent with the earlier fluence-MCNP5 work performed by Dugas *et al.*[4] If the scattering cross section used for the Dugas work is increased by 13%, as discussed in Section I, the average % difference reported by Dugas increases from -6.4 ± 0.8% to 5.8 ± 0.8%, which is consistent with the 35 keV results presented in this work.

**IV. CONCLUSIONS**

This work is important for verifying the AAPM TG-61 ion chamber dosimetry used to calibrate dose output from monochromatic x-ray beams, which have been used for photoactivated Auger electron therapy. Two significant improvements were made to the method used to determine the fluence-MCNP5 dose distribution described by Dugas *et al.*[4]: (1) Compton cross section calculations were revised to include electron binding effects, and (2) the Rayleigh scatter contribution was incorporated into the fluence calculations. In addition, the fluence contribution from n=2 photons was included in the final dose distribution, although this has a relatively small effect due to their low number. The results show that the TG-61 ion-chamber dosimetry agree with the fluence-MCNP5 dosimetry to within approximately 7% and 3% at beam energies of 25 and 35 keV, respectively, for PMMA depths of 0.6 to 7.7 cm. This is an acceptable level of agreement for ongoing cell irradiation dosimetry. Resolution of the differences in the two dose methods might benefit from the use of an additional dose measurement device, i.e. a calorimeter, and an extension of TG-61 to include monochromatic x-ray beams.

## ACKNOWLEDGEMENTS

This research is supported by contract W81XWH-10-1-0005 awarded by The U.S Army Research Acquisition Activity, 820 Chandler Street, Fort Detrick, MD 21702-5014. This paper does not necessarily reflect the position or policy of the Government, and no official endorsement should be inferred.
## REFERENCES

[1] J. P. Dugas, M. E. Varnes, E. Sajo, C. E. Welch, K. Ham, and K. R. Hogstrom, "Dependence of cell survival on iododeoxyuridine concentration in 35-keV photon-activated Auger electron radiotherapy", Int. J. Radiation Oncology Biol. Phys. **79**, 255-261 (2011).

[2] S. D. Oves, K. R. Hogstrom, K. Ham, E. Sajo, J. P. Dugas, "Dosimetry intercomparison using a 35-keV X-ray synchrotron beam", Eur. J. Radiol. **68S**, 121-125 (2008).

[3] C.-M. Ma, C. W. Coffey, L. A. DeWerd, C. Liu, R. Nath, S. M. Seltzer, J. P. Seuntjens, "AAPM protocol for 40-300 kV x-ray beam dosimetry in radiotherapy and radiobiology", Med. Phys. **28**, 868-893 (2001).

[4] J. P. Dugas, S. D. Oves, E. Sajo, K. L. Matthews, K. Ham, K. R. Hogstrom, "Monochromatic beam characterization for Auger electron dosimetry and radiotherapy", Eur. J. Radiol. **68S**, 137-141 (2008).

[5] J. H. Hubbell, Wm. J. Veigele, E. A. Briggs, R. T. Brown, D. T. Cromer, R. J. Howerton, "Atomic Form Factors, Incoherent Scattering Functions, and Photon Scattering Cross Sections", J. Phys. Chem. Ref. Data **4,** 471-538 (1975).

[6] D. Cullen, J. Hubbell, L. Kissel, "EPDL97: the evaluated photon library", Technical Report UCRL-50400 Vol. 6 Rev. 5 (Lawrence Livermore National Laboratory Report, 1997).

[7] T. Schoonjans, A. Brunetti, B. Golosio, M. Sanchez del Rio, V. A. Solé, C. Ferrero, L. Vincze, "The xraylib library for X-ray-matter interactions. Recent developments", Spectrochim. Acta Part B **66**, 776-784 (2011).

[8] J. H. Hubbell, S. M Seltzer, "Tables of X-ray Mass Attenuation Coefficients and Mass Energy-Absorption Coefficients from 1 keV to 20 MeV for Elements Z = 1 to 92 and 48 Additional Substances of Dosimetric Interest", version 1.4 (National Institute of Standards and Technology, Gaithersburg, MD, 2004) (available URL: http://www.nist.gov/pml/data/xraycoef/index.cfm).

[9] G. F. Knoll, "Radiation Detection and Measurement", 3rd ed. (Wiley, Hoboken, NJ, 2000).
**Medical Physics, Vol. 39, No. 12, p. 7462-7469, December 2012**                                                                                                              **22**